# Lack of Self-affinity and Anomalous Roughening in Growth Processes

Juan M. López[1,2] and Miguel A. Rodríguez[1]

[1] *Instituto de Física de Cantabria (CSIC-UC), Avenida de Los Castros, E-39005 Santander, Spain.*

[2] *Departamento de Física Moderna, Universidad de Cantabria, E-39005 Santander, Spain.*

We contrast analytical results of a variety of growth models involving subdiffusion, thermal noise and quenched disorder with simulations of these models, concluding that the assumed self-affinity property is more an exception than a rule. In our two dimensional models, self-affine surfaces may only appear when the roughness exponent is $\chi = 1/2$ or $\chi = 1$. A new scaling picture, which leads to more suitable ways of determining the scaling exponents, is proposed when lack of self-affinity exists.

5.40.+j, 05.70.Ln, 68.35.Fx

Interface roughening observed in many growth phenomena [1,2] can be explained as produced by the interplay between a surface tension, which tends to smooth the interfacial curvature, and some kind of noise coming either from thermal fluctuations or spatial inhomogeneities (*i.e.* quenched disorder). In the simplest model exhibiting kinetic roughening, the Edwards-Wilkinson (EW) model [3], both effects are involved by means of a diffusion term and an additive thermal noise. In order to introduce some symmetries existing in many real systems, a nonlinear term must be added to the EW model and this gives rise to the Kardar-Parisi-Zhang (KPZ) equation [4]. The two-dimensional KPZ model successfully describes a remarkable diversity of growth processes and is an excellent example of universality [1,2]. Simulations and analytical calculations indicate that the two-dimensional KPZ model leads to self-affine interfaces with simple scaling laws. One of the most important quantities used to characterize the scaling of the interface is the global width:

$$\sigma(L,t) = \{(\langle h^2(x,t)\rangle - \langle h(x,t)\rangle^2)^{1/2}\}, \qquad (1)$$

where $L$ is the system size, $<..>$ denotes a spatial average over the whole system and $\{..\}$ over realizations. For the early times regime, the interface width is expected to be of the form $\sigma(L,t) \sim t^\beta$ where $\beta$ is called the time exponent. At a characteristic saturation time $t_s(L) \sim L^z$, where $z$ is the dynamic exponent, the horizontal correlation length reaches the system size and the width saturates, $\sigma(L, t > t_s) \sim L^\chi$, with a roughness exponent $\chi$. This is the so-called Family-Vicsek scaling *ansatz* [1,2].

Thus, the roughness exponent is calculated from the scaling of the width in saturation. To measure $\chi$ in an experiment or simulation, one has to consider systems with different sizes and wait for saturation at times $t_s(L) \sim L^z$. This way of determining $\chi$ is not very convenient from a computational point of view due to the large system sizes in which simulations must be done. The situation in experiments is not better, because one should perform measurements in samples with different sizes. For that, in practice an alternative method is usually used. The width is evaluated over a window of size $l \ll L$ and one waits for saturation a shorten time $t_s(l) \ll t_s(L)$. This local width is believed to scale as the global width, $\sigma(l,t) \sim t^\beta$ for short times, and the roughness exponent is obtained from the scaling of $\sigma(l,t)$ at times $t \gg l^z$:

$$\sigma(l, t \gg l^z) \sim l^{\chi_{loc}}. \qquad (2)$$

Equivalently, the height difference correlation function, $G(l,t) = \{\langle (h(l+x,t) - h(x,t))^2 \rangle\}$, can be used to determine the scaling exponents because $G(l,t) \sim \sigma^2(l,t)$. It is a general belief [1,2] that the roughness exponent $\chi_{loc}$ obtained in this way is equal to the exponent $\chi$ that one gets from the global width. In fact, this is correct when the interface is self-affine.

In general, this picture has been extrapolated to other kinds of growth phenomena involving higher dimensions (KPZ in an arbitrary dimension) [1,2], subdiffusive behavior in epitaxial growth phenomena [5,2] and quenched noise in porous media [2,6–10]. In contrast to the two-dimensional KPZ case, now experiments and simulations show a great discrepancy. There are numerical evidences about an *anomalous* scaling behavior in growth processes in which the roughness exponent is larger than one ($\chi > 1$). In numerical studies of models related to the molecular-beam epitaxy (MBE) growth, it has been found [11,12] that global roughness (measured by the interface width) gives different scaling exponents than local roughness (measured by either a height difference correlation function or the local width in a window of size $l$). Similar results have been numerically obtained in the case of the EW equation with quenched noise [8–10]. This equation has attracted much attention in recent years [2,6–10] because it is the simplest way of describing driven interfaces in quenched disordered media. Surprisingly, it seems that local and global dynamic scaling behaviors are distinctly different for some growth processes. This anomalous scaling behavior is not predicted by the conventional dynamic scaling hypothesis of Family-Vicsek and it is not well understood.

The most recent interpretation [8–11,13,14] of this anomalous scaling is that the conventional dynamic scaling fails in the case of roughness exponent larger than one, $\chi > 1$. In the case of MBE growth some other effects as the breaking of spectra [15] and the existence of a groove instability [11] have been also associated with anomalous scaling. However, very recently, Das Sarma *et. al.* [16] have studied MBE growth models with a global roughness $\chi < 1$ in which an anomalous scaling behavior appeared as well. So, it seems that a roughness $\chi > 1$ is not the only reason for anomalous scaling behavior. Moreover, the situation is further complicated due to fact that for most of these growth models there is not an analytical value of the roughness exponent.

Our aim in this Letter is to show that anomalous scaling occurs not only whenever $\chi > 1$ but also in growth processes in which $\chi < 1$ different roughness exponents may appear at small and large length scales. As a consequence the interface is not self-affine and a new scaling picture emerges. Our conclusions are based on analytical solutions (and comparison with simulation) of growth models with subdiffusive behavior and/or quenched disorder.

*Models with uniform diffusion.-* The inequality $\chi \neq \chi_{loc}$ has been numerically found in models without analytical solutions. In the following we are going to consider models with analytical solutions that will allow us to understand the existence of a local exponent $\chi_{loc}$. Let us start by studying models with diffusive or subdiffusive behavior and thermal noise governed by the equation:

$$\frac{\partial h(x,t)}{\partial t} = (-1)^{m+1} \frac{\partial^{2m} h(x,t)}{\partial x^{2m}} + \eta(x,t), \tag{3}$$

where $\eta(x,t)$ is a Gaussian noise with correlation:

$$\langle \eta(x,t)\eta(x',t') \rangle = \Delta_a(x-x')\delta(t-t'),$$

the correlator $\Delta_a(x)$ is a normalized function decaying rapidly to zero over a finite distance $a$. The case $m = 1$ corresponds to the EW model [3], for $m = 2$ we have the linear model [5,16,11,13] used in MBE growth and in general by increasing $m$ more subdiffusive processes are obtained.

Since these models are linear the scaling exponents can be easily calculated: $\chi = (2m-1)/2$, $\beta = (2m-1)/4m$ and $z = 2m$. We have numerically solved Eq.(3) for $m = 1$ and $m = 2$. In all cases there exists a good agreement between the analytically calculated $\chi$ and simulation results when the Eq.(1) was used. For $m = 1$ also the local method, Eq.(2), gives the correct exponent $\chi_{loc} = \chi = 1/2$. However, for $m = 2$ there is a remarkable discrepancy between the local roughness exponent obtained in our simulation, $0.83 \pm 0.05$, and the theoretical value $\chi = 3/2$. This discrepancy indicates that the interface is not self-affine and different roughness exponents exist at large and small length scales. The analytical treatment of (3) allows to understand the existence of this anomalous scaling. Consider the height difference correlation function $G(l,t)$ under the $l/L \ll 1$ condition:

$$G(l,t) = t^{2\beta} \int_{-\infty}^{\infty} du \frac{(1-e^{-2u^{2m}})}{u^{2m}} (1-e^{-iul/t^{1/z}}) \Delta_{a/t^{1/z}}(u) \tag{4}$$

$\Delta_a(k)$ being the Fourier transform of the correlator $\Delta_a(l)$. For early times, $l/t^{1/z} \gg 1$, we have $G(l,t) \sim c(t)t^{2\beta}$, $c(t)$ slowly varying and in the long time limit, $l/t^{1/z} \ll 1$, $G(l,t) \sim t^{2\beta}(l/t^{1/z})b(l/a)$ (where $b(l/a) \sim constant$) when $m = 1$ and $G(l,t) \sim t^{2\beta}(l/t^{1/z})^2$ when $m \geq 2$. Then for small length scales, $l \ll t^{1/z}$, we have the general scaling

$$G(l,t) \sim t^{2\beta}(l/t^{1/z})^n \sim t^{2\beta_*} l^{2\chi_{loc}}, \tag{5}$$

where $\beta_* = \beta - n/2z$, $\chi_{loc} = n/2$, and the integer $n = 1$ or $n = 2$ depending on the model. Note that this scaling would occur whenever $G(l,t) \sim t^{2\beta} g(l/t^{1/z}, a/t^{1/z})$, where the scaling function $g(x,y)$ is analytical in $x = 0$ and $g(0,0) = 0$. For true self-affine interfaces the exponent $\beta_* = 0$ and the Family-Vicsek scaling is recovered. This only occurs for $\beta z = n/2$, in such a way that the interface is self-affine whenever $\chi = n/2$ and in this case $\chi = \chi_{loc}$. Thus, Eq.(5) gives the self-affinity condition in the sense that whenever $\beta_* \neq 0$ the interface is not self-affine.

From Eq.(5) for $m = 1$ (case with $n = 1$), we obtain that $G(l,t)$ becomes constant and independent of time, $G(l, t \gg l^z) \sim l$, as corresponds to the usual Family-Vicsek scaling picture. However, for $m \geq 2$ ($n = 2$ in this case) for small length scales, $l \ll t^{1/z}$, we have $G(l,t) \sim l^2 t^{1-3/2m}$, and there is not saturation of $G(l, t \gg l^z)$ at long times. Contrary to what is expected in the usual scaling, the height difference correlation (or equivalently the square local width) scales with time as $t^{2\beta_*}$ and a new temporal exponent appears, $\beta_* = \beta - 1/z = (2m-3)/4m$ describing the long time behavior of the local width. In particular, the linear model of MBE [5,16,11,13] (Eq. (3) for $m = 2$) exhibits an anomalous scaling in the sense that in the earliest time regime the local width (measured by the quantity $G(l,t)$) scales as $t^{2\beta} \sim t^{3/4}$ and does not saturate but $G(l,t) \sim l^2 t^{2\beta_*}$ for $l \ll t^{1/z} \ll L$ (where $\beta_* = \beta - 1/z = 1/8$). Our numerical results (see Fig. 1) for Eq.(3) with $m = 2$ are in excellent agreement with this analysis. The saturation regime in this case is dominated by the saturation of the largest scales $l \sim L$. At time $t \sim L^z$ saturation occurs in the whole system and $G(l, t \to \infty) \sim l^2 L^{2\beta_* z} \sim l^2 L^{2\chi-2}$ for $l \ll L \ll t^{1/z}$. When saturation occurs, $G(l, t \to \infty)$ exhibits two different spatial scaling behaviors. For $l \ll L$, $G(l, t \to \infty)$ scales with the window size as $l^2$ (which corresponds to a local roughness exponent $\chi_{loc} = 1$) and for large scales, $l \sim L$, $G(L, t \to \infty) \sim \sigma^2(L, t \to \infty) \sim L^{2\chi}$. This crossover behavior leads to an *effective* exponent $\chi_{eff}$ in between large and short length scales, that is the scaling behavior measured in simulations (in our simulation we found $\chi_{eff} = 0.83 \pm 0.05$). The local exponent $\chi_{eff}$ apparently should take an intermediate value between the interpolated exponents $\chi_{loc} = 1$ and $\chi = 3/2$ but it is never the case because $G(l,t)$ is actually greater than the corresponding value if the interface were saturated and then the interpolation to largest scales occurs with decreasing slopes. As a consequence any determination of the local roughness exponent from Eq.(2) gives underestimated values of $\chi_{loc}$.

*Models with random diffusion.–* As we have discussed above, models governed by Eq.(3) give anomalous scaling for $m \geq 2$. In this case the roughness is always $\chi = (2m-1)/2 > 1$ and $\chi_{loc} \sim 1$ on short length scales. The question that we would like to discuss now is if the lack of self-affinity is due to a roughness exponent $\chi > 1$ or, on the contrary, anomalous scaling can exist even in growth models with $\chi < 1$. The most extended opinion [8–11,13,14] is that interfaces with $\chi > 1$ are not properly fractal objects and self-affinity is not a sure feature. But self-affinity for interfaces in which $\chi < 1$ is a simple hypothesis that, as we will show, in our models is only valid for $\chi = 1/2$ or 1. This can be readily verified in a kind of growth model in which $\chi$ takes a continuous range of values controlled by the intensity of disorder in the diffusion coefficient. This model can be used to describe a fluid interface advancing through stratified porous materials [17] with long range correlations in the growth direction. The growth model that we study here is given by

$$\frac{\partial h(x,t)}{\partial t} = \frac{\partial}{\partial x} D(x) \frac{\partial}{\partial x} h(x,t) + \eta(x,t). \qquad (6)$$

$D(x) > 0$ is the columnar diffusivity coefficient that controls the interconnectivity between neighbor channels. This equation can be viewed like (3) for $m = 1$ but with a quenched random diffusion term. A similar model was recently studied in Ref. [17].

It is known [17] that Eq.(6) can be considered as a model of diffusing particles with time-dependent sources and sinks in such a way that it is possible using the well known theory of transport through random media to take the usual stochastic description for $D(x)$. So, $D(x)$ is taken to be an uncorrelated random field and distributed according to a probability density $P(D) = N_\alpha D^{-\alpha} f_c(D/D_{max})$, where $N_\alpha$ is a normalization constant and $f_c$ a cutoff function. The parameter $\alpha$ characterizes the intensity of disorder. Disorder is called *weak* for $\alpha = 0$. *Strong* disorder occurs for $0 < \alpha < 1$ because of the divergence of $P(D)$ at origin. The model so formulated is easy to simulate numerically and allows analytical treatment. As it has been shown [17], the inhomogeneous problem (6) is statistically equivalent to an homogeneous one with a time dependent diffusion coefficient given by the so-called Effective Medium Approximation condition [17]. With the above defined $P(D)$ the equivalent diffusion length scales as $l_D(t) \sim t^{(1-\alpha)/(2-\alpha)}$ and consequently the dynamical exponent $z$ will be given by $z = (2-\alpha)/(1-\alpha)$. The exponents $\beta$ and $\chi$ can be easily calculated, obtaining $\beta = 1/2(2-\alpha)$ and $\chi = 1/2(1-\alpha)$. Note that, as we have mentioned above, a continuous range of roughness exponents (from $\chi = 1/2$ to $\chi = \infty$) are obtained by varying the intensity of disorder $\alpha$.

The height difference correlation function can be calculated [17] in the asymptotic time regime from (6) and is given by :

$$G(l,t) \sim t^{2\beta} \int_{-\infty}^{\infty} du \int_0^1 dv \int ds \int_0^s ds\prime \frac{e^{sv}}{(s-s\prime-D(s-s\prime)u^2)(s\prime-D(s\prime)u^2)}(1-e^{-iul/t^{1/z}})\Delta_{a/t^{1/z}}(u), \qquad (7)$$

where $D(s) \sim s^{\alpha/(2-\alpha)}$ is the Laplace transform of the time-dependent diffusion coefficient. Since (6) has the same analytical behavior as (3) for $m = 1$ (with an equivalent time-dependent diffusion coefficient $D(t)$) it is easy to see that the scaling of $G(l,t)$, Eq.(7), for small scales $l \ll t^{1/z}$ is also given by Eq.(5) with $n = 1$ (with the corresponding $\alpha$-dependent exponents).

For small length scales, $l \ll t^{1/z} \ll L$, from Eq.(5) we have that the square local width $G(l,t)$ crosses over to a different temporal regime $G(l, t \gg l^z) \sim t^{2\beta_*} l$ where the time exponent is $\beta_* = \beta - 1/2z = \alpha/2(2-\alpha)$. Thus, the interface would be self-affine only in the case of weak disorder $\alpha = 0$ (*i.e.* the EW equation). For $0 < \alpha < 1$, the interface is not a self-affine object and an anomalous scaling occurs. In Fig.2 our numerical results for the local width versus time are presented for three degrees of disorder. We can see that the exponent $\beta_*$ predicted by our scaling is in very good agreement with simulation for all $\alpha$ (see Table I).

In true saturation, at very long times, $l \ll L \ll t^{1/z}$, the scaling behavior of $G(l,t)$ can be obtained from (5) for $n=1$ and we have $G(l, t \to \infty) \sim lL^{2\chi-1}$. So, in the case of strong disorder a local roughness $\chi_{loc} = 1/2$ is obtained. Certainly, as we can see in Table I, the crossover to the global roughness $\chi = 1/[2(2-\alpha)] > 1/2$ for larger scales yields lower numerical values of the local roughness exponent. In all cases, included those with $\chi < 1$, the interface is not self-affine as we claimed above. From our models it is clear that the standard exponents $\chi$, $\beta$, and $z$ can not characterize completely the scaling behavior of a growing interface. For instance, note that the linear MBE model, Eq.(3) for $m=2$, leads to the same critical exponents $\beta$, $\alpha$ and $z$, as the model with random diffusion, Eq.(6), in the case of a disorder parameter $\alpha = 2/3$, however both models have a very different local roughness exponent in such a way that they belong to distinct universality classes.

In conclusion, we have dealt with a variety of growth models that allowed us to an analytical treatment. Analytical results have been compared to simulations focusing our interest on the scaling of the local width at small length scales, $l \ll L$. We have concluded that, except for growth with $\chi = n/2$ (where $n = 1, 2$), interfaces in two dimensions are not self-affine and the Family-Vicsek dynamic scaling fails. In addition, we have seen that a new time exponent $\beta_* = \beta - n/2z$ (where $n = 1, 2$ depending on the model) has to be introduced in order to get a complete description of the scaling behavior of the interface. Moreover, the roughness exponent $\chi_{eff}$, commonly obtained in a number of simulations (and experiments), is an effective exponent affected by a crossover from the local scaling exponent $\chi_{loc}$ to the global roughness exponent $\chi$. This effective roughness exponent does not give any information about the value of the scaling exponents $\chi$, $\beta$ and $z$, however, $\chi_{eff}$ is important from a physical point of view to obtain the effective roughness measured in experiments and to distinguish the corresponding universality class.

Although the new scaling picture, proposed in Eq.(5), has been obtained from the study of particular models, we believe that it can be applied to more general growth processes (in two dimensions). The scaling (5) is based on property of the scaling function, $g$, of being analytic at small length scales and this is expected to be verified for a broad class of growth problems. The existence of kinetic roughening satisfying (5) but with values of $n \neq 1, 2$ must be investigated further since, in this case, $\chi_{loc}$ would be a new and interesting scaling exponent [18].

This work has been supported by DGICyT of the Spanish Government, Project No. PB93–0054–C02–02.

[1] F.Family T.Vicsek eds., *Dynamics of Fractal Surfaces* (World Scientific, Singapore, 1991).
[2] T. Halpin–Healey and Y.-C. Zhang, Phys. Rep. **254**, 215 (1995); A.-L. Barabasi and H.E. Stanley, *Fractal concepts in surface growth*, (Cambridge University Press, Cambridge, 1995).
[3] S.F. Edwards and D.R. Wilkinson, Proc. R. Soc. Lond. A**381**, 17 (1982).
[4] M. Kardar, G. Parisi and Y.-C. Zhang, Phys. Rev. Lett. **56**, 889 (1986).
[5] Z. W. Lai and S. Das Sarma, Phys. Rev. Lett. **66**, 2348 (1991).
[6] J.M. López, Phys. Rev. E **52**, R1296 (1995).
[7] J.M. López, M.A. Rodríguez, A. Hernández-Machado, and A. Díaz-Guilera, Europhys. Lett., **29**, 197, (1995)
[8] S. Roux and A. Hansen, J. Phys. I (France) **4**, 515 (1994).
[9] H.J. Jensen, J. Phys. A **28**, 1861 (1995).
[10] L.A.N. Amaral, A.L. Barabasi, H.A. Makse and H.E. Stanley, Phys. Rev. E **52**, 4087 (1995)
[11] J.G. Amar, P. M. Lam, F. Family, Phys. Rev. E **47**, 3242 (1993).
[12] M. Schroeder, M. Siegert, D.E. Wolf, J.D. Shore, and M. Plischke, Europhys. Lett. **24**, 563 (1993).
[13] S. Das Sarma, S.V. Ghaisas, and J.M. Kim, Phys. Rev E **49**, 122 (1994).
[14] H. Leschhorn and L.-H. Tang, Phys. Rev. Lett. **70**, 2973 (1993)
[15] M. Siegert and M. Plischke, Phys. Rev. Lett. **68**, 2035 (1992) .
[16] S. Das Sarma, C.J. Lanczycki, R. Kotlyar, and V. Ghaisas, Phys. Rev. E **53**, 359 (1996).
[17] J.M. López and M.A. Rodríguez, Phys. Rev. **52** E, 6442 (1995).
[18] In Ref. [16] several MBE growth models were simulated obtaining a value of $\chi_{loc} \sim 0.6 \pm 0.1$ ($\alpha'$ in the quoted paper). These models would belong to the class of models satisfying (5) with $n = 1$.

TABLE I. Scaling exponents for the model with random diffusion, Eq.(6), for several degrees of disorder. Theoretical values of $\beta$, $\alpha$, and $\beta_*$ are given in the text. The effective roughness exponent is always lower than the local roughness $\chi_{loc} = 1/2$. Note that $\chi_{loc}$ remains roughly constant despite the great difference existing in $\chi$

| $\alpha$ | $\beta$ (exact) | $\beta$ (simulation) | $\chi$ (exact) | $\chi_{eff}$ (simulation) | $\beta_*$ (theory) | $\beta_*$ (simulation) |
|---|---|---|---|---|---|---|
| 2/3 | 3/8 | $0.38 \pm 0.05$ | 3/2 | $0.35 \pm 0.05$ | 1/4 | $0.27 \pm 0.03$ |
| 1/2 | 1/3 | $0.34 \pm 0.01$ | 1 | $0.39 \pm 0.04$ | 1/6 | $0.18 \pm 0.03$ |
| 1/3 | 3/10 | $0.31 \pm 0.02$ | 3/4 | $0.45 \pm 0.05$ | 1/10 | $0.12 \pm 0.04$ |

FIG. 1. Square local width vs time for the linar MBE model, $m = 2$, calculated over a window size $l = 5$ (squares), $l = 10$ (triangles), and $l = 150$ (crosses). The continuous line corresponds to the fit of the square global width, $\sigma^2(L, t) \sim t^{2\beta}$. The dashed line is the adjusted curve of the data for $l = 5$ and its slope $0.26 \pm 0.02$ corresponds to $2\beta_*$. A system size of $L = 1000$ was used and an average over 15 realizations of disorder was performed.

FIG. 2. Square local width vs time for the random diffusion model, Eq.(6), averaged over a length scale $l = 5$ (squares), $l = 10$ (triangles), and $l = 150$ (crosses) for different degrees of disorder. A system size $L = 1000$ and 15 realizations of disorder were used. As in Fig.1, the continuous line slopes correspond to the values of $2\beta$ obtained in simulation. The dashed line is the adjusted curve of the data for $l = 5$. A comparison with our scaling theory is given in Table I.

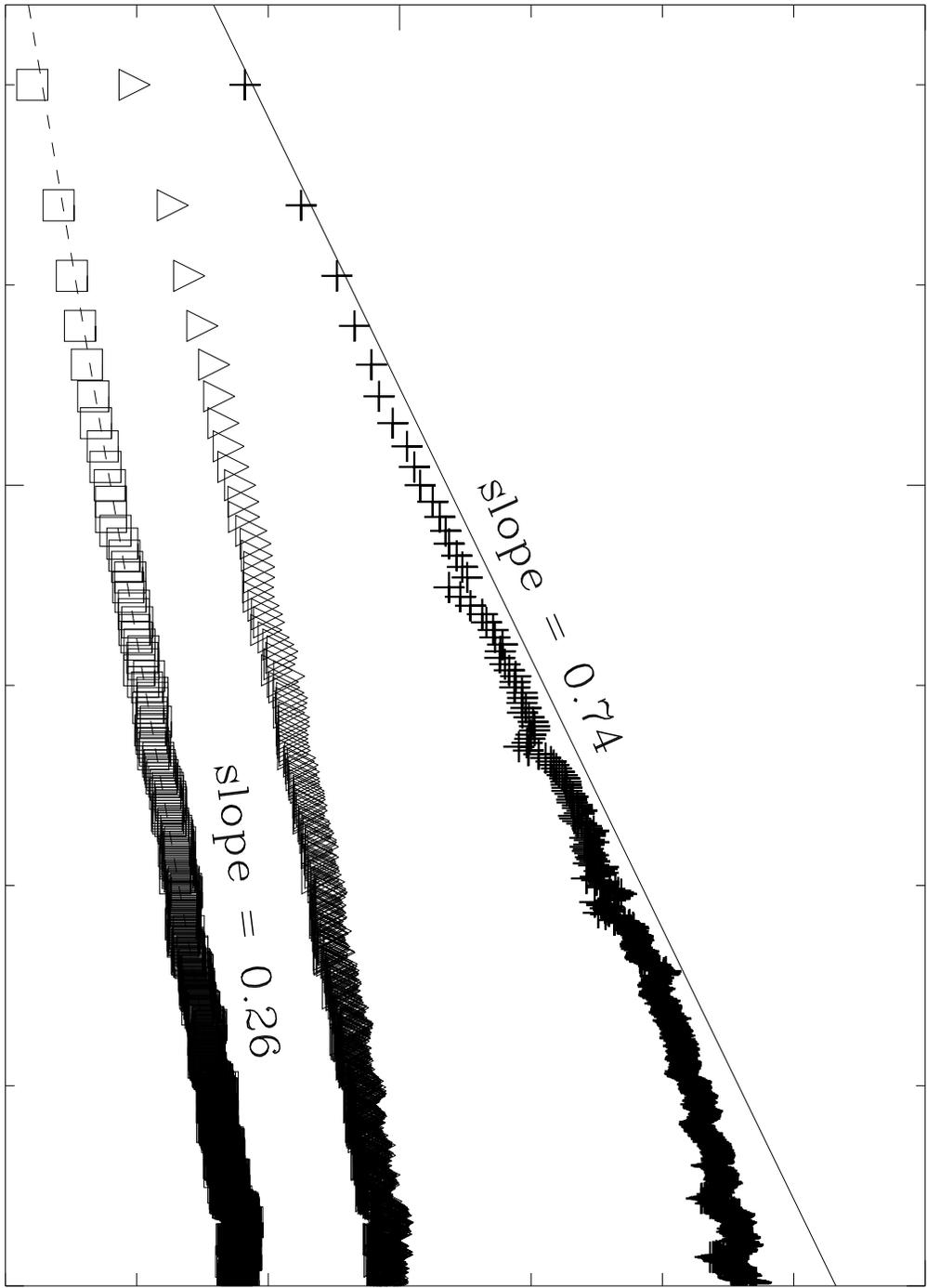

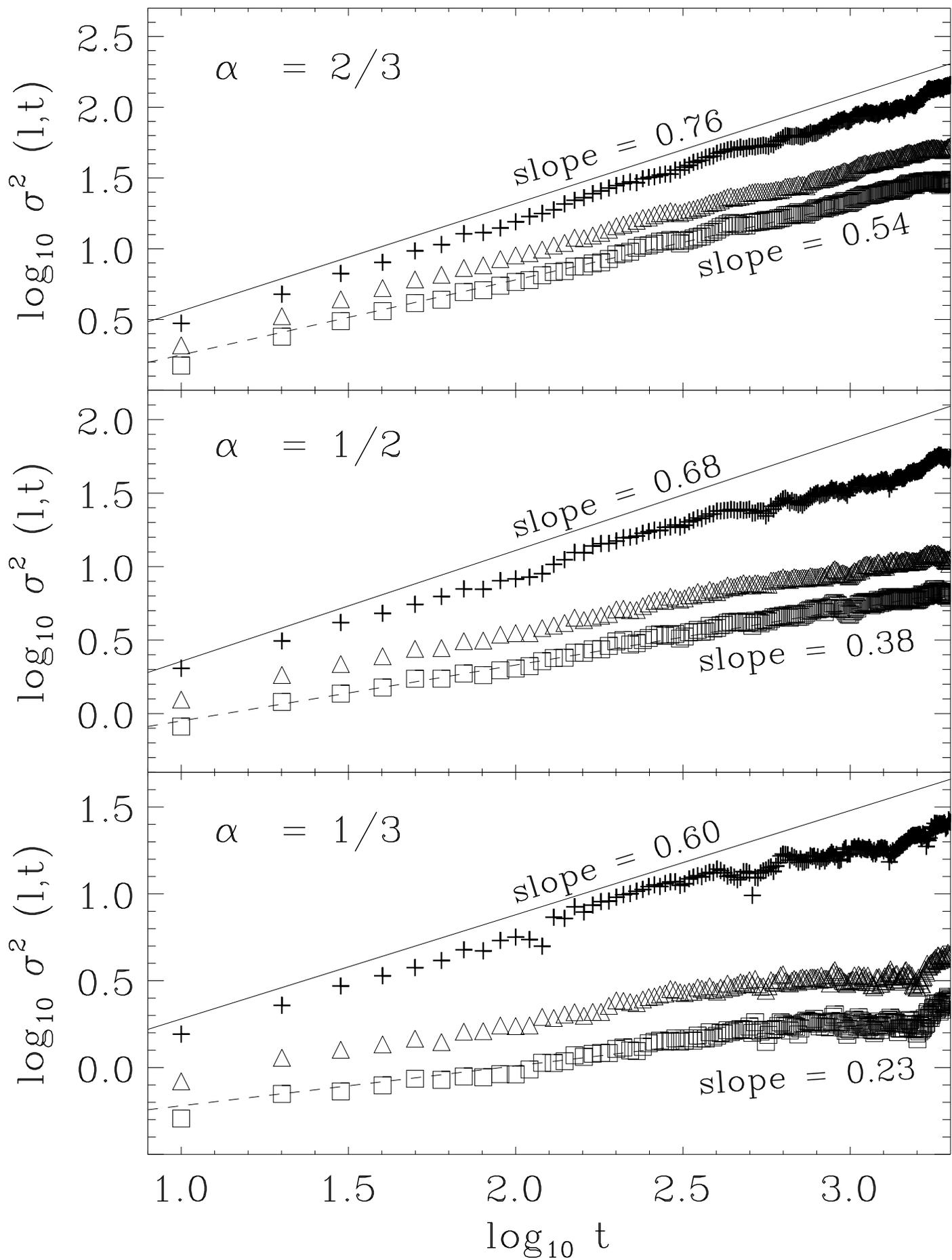